\input harvmac

\def\ev#1{\langle#1\rangle}

\input amssym
\input epsf.tex

\newcount\figno
\figno=0
\def\fig#1#2#3{
\par\begingroup\parindent=0pt\leftskip=1cm\rightskip=1cm\parindent=0pt
\baselineskip=11pt
\global\advance\figno by 1
\midinsert
\epsfxsize=#3
\centerline{\epsfbox{#2}}
\vskip 12pt
{\bf Fig. \the\figno:} #1\par
\endinsert\endgroup\par
}
\def\figlabel#1{\xdef#1{\the\figno}}
\def\encadremath#1{\vbox{\hrule\hbox{\vrule\kern8pt\vbox{\kern8pt
\hbox{$\displaystyle #1$}\kern8pt}
\kern8pt\vrule}\hrule}}


\newfam\frakfam
\font\teneufm=eufm10
\font\seveneufm=eufm7
\font\fiveeufm=eufm5
\textfont\frakfam=\teneufm
\scriptfont\frakfam=\seveneufm
\scriptscriptfont\frakfam=\fiveeufm


\def\bb{
\font\tenmsb=msbm10
\font\sevenmsb=msbm7
\font\fivemsb=msbm5
\textfont1=\tenmsb
\scriptfont1=\sevenmsb
\scriptscriptfont1=\fivemsb
}



\newfam\dsromfam
\font\tendsrom=dsrom10
\textfont\dsromfam=\tendsrom
\def\ds{\fam\dsromfam \tendsrom}


\newfam\mbffam
\font\tenmbf=cmmib10
\font\sevenmbf=cmmib7
\font\fivembf=cmmib5
\textfont\mbffam=\tenmbf
\scriptfont\mbffam=\sevenmbf
\scriptscriptfont\mbffam=\fivembf


\newfam\mbfcalfam
\font\tenmbfcal=cmbsy10
\font\sevenmbfcal=cmbsy7
\font\fivembfcal=cmbsy5
\textfont\mbfcalfam=\tenmbfcal
\scriptfont\mbfcalfam=\sevenmbfcal
\scriptscriptfont\mbfcalfam=\fivembfcal


\newfam\mscrfam
\font\tenmscr=rsfs10
\font\sevenmscr=rsfs7
\font\fivemscr=rsfs5
\textfont\mscrfam=\tenmscr
\scriptfont\mscrfam=\sevenmscr
\scriptscriptfont\mscrfam=\fivemscr




\def\bar{\overline}
\def\b{\bar}
\def\bsq#1{{{\b{#1}}^{\lower 2.5pt\hbox{$\scriptstyle 2$}}}}
\def\bexp#1#2{{{\b{#1}}^{\lower 2.5pt\hbox{$\scriptstyle #2$}}}}
\def\dotexp#1#2{{{#1}^{\lower 2.5pt\hbox{$\scriptstyle #2$}}}}


\def\rt2{\sqrt{2}}
\def\half {{1 \over 2}}

\def\tr{\mathop{\rm tr}}


\font\tenbifull=cmmib10
\font\tenbimed=cmmib7
\font\tenbismall=cmmib5
\textfont9=\tenbifull \scriptfont9=\tenbimed
\scriptscriptfont9=\tenbismall

\mathchardef\bbGamma="7000
\mathchardef\bbDelta="7001
\mathchardef\bbPhi="7002
\mathchardef\bbAlpha="7003
\mathchardef\bbXi="7004
\mathchardef\bbPi="7005
\mathchardef\bbSigma="7006
\mathchardef\bbUpsilon="7007
\mathchardef\bbTheta="7008
\mathchardef\bbPsi="7009
\mathchardef\bbOmega="700A
\mathchardef\bbalpha="710B
\mathchardef\bbbeta="710C
\mathchardef\bbgamma="710D
\mathchardef\bbdelta="710E
\mathchardef\bbepsilon="710F
\mathchardef\bbzeta="7110
\mathchardef\bbeta="7111
\mathchardef\bbtheta="7112
\mathchardef\bbiota="7113
\mathchardef\bbkappa="7114
\mathchardef\bblambda="7115
\mathchardef\bbmu="7116
\mathchardef\bbnu="7117
\mathchardef\bbxi="7118
\mathchardef\bbpi="7119
\mathchardef\bbrho="711A
\mathchardef\bbsigma="711B
\mathchardef\bbtau="711C
\mathchardef\bbupsilon="711D
\mathchardef\bbphi="711E
\mathchardef\bbchi="711F
\mathchardef\bbpsi="7120
\mathchardef\bbomega="7121
\mathchardef\bbvarepsilon="7122
\mathchardef\bbvartheta="7123
\mathchardef\bbvarpi="7124
\mathchardef\bbvarrho="7125
\mathchardef\bbvarsigma="7126
\mathchardef\bbvarphi="7127






\def\1{{\ds 1}}

\def\Z{\hbox{$\bb Z$}}


\noblackbox

\def\unit{\relax{\rm 1\kern-.26em I}}
\def\nada{\relax{\rm 0\kern-.30em l}}


\noblackbox
\def\IL{\relax{\rm I\kern-.18em L}}
\def\IH{\relax{\rm I\kern-.18em H}}
\def\IR{\relax{\rm I\kern-.18em R}}
\def\IC{\relax\hbox{$\inbar\kern-.3em{\rm C}$}}
\def\IZ{\relax\ifmmode\mathchoice
{\hbox{\cmss Z\kern-.4em Z}}{\hbox{\cmss Z\kern-.4em Z}} {\lower.9pt\hbox{\cmsss Z\kern-.4em Z}}
{\lower1.2pt\hbox{\cmsss Z\kern-.4em Z}}\else{\cmss Z\kern-.4em Z}\fi}

\def\partialslash{\not{\hbox{\kern-2pt $\partial$}}}


\font\manual=manfnt \def\dbend{\lower3.5pt\hbox{\manual\char127}}

\def\IZ{\relax\ifmmode\mathchoice
{\hbox{\cmss Z\kern-.4em Z}}{\hbox{\cmss Z\kern-.4em Z}} {\lower.9pt\hbox{\cmsss Z\kern-.4em Z}}
{\lower1.2pt\hbox{\cmsss Z\kern-.4em Z}}\else{\cmss Z\kern-.4em Z}\fi}
\def\half {{1\over 2}}

\def\bar{\overline}

\def\rt2{\sqrt{2}}
\def\irt2{{1\over\sqrt{2}}}

\def\slashchar#1{\setbox0=\hbox{$#1$}           
   \dimen0=\wd0                                 
   \setbox1=\hbox{/} \dimen1=\wd1               
   \ifdim\dimen0>\dimen1                        
      \rlap{\hbox to \dimen0{\hfil/\hfil}}      
      #1                                        
   \else                                        
      \rlap{\hbox to \dimen1{\hfil$#1$\hfil}}   
      /                                         
   \fi}

\def\foursqr#1#2{{\vcenter{\vbox{
    \hrule height.#2pt
    \hbox{\vrule width.#2pt height#1pt \kern#1pt
    \vrule width.#2pt}
    \hrule height.#2pt
    \hrule height.#2pt
    \hbox{\vrule width.#2pt height#1pt \kern#1pt
    \vrule width.#2pt}
    \hrule height.#2pt
        \hrule height.#2pt
    \hbox{\vrule width.#2pt height#1pt \kern#1pt
    \vrule width.#2pt}
    \hrule height.#2pt
        \hrule height.#2pt
    \hbox{\vrule width.#2pt height#1pt \kern#1pt
    \vrule width.#2pt}
    \hrule height.#2pt}}}}
\def\psqr#1#2{{\vcenter{\vbox{\hrule height.#2pt
    \hbox{\vrule width.#2pt height#1pt \kern#1pt
    \vrule width.#2pt}
    \hrule height.#2pt \hrule height.#2pt
    \hbox{\vrule width.#2pt height#1pt \kern#1pt
    \vrule width.#2pt}
    \hrule height.#2pt}}}}
\def\sqr#1#2{{\vcenter{\vbox{\hrule height.#2pt
    \hbox{\vrule width.#2pt height#1pt \kern#1pt
    \vrule width.#2pt}
    \hrule height.#2pt}}}}

\def\figin{\epsfcheck\figin}\def\figins{\epsfcheck\figins}
\def\epsfcheck{\ifx\epsfbox\UnDeFiNeD
\message{(NO epsf.tex, FIGURES WILL BE IGNORED)}
\gdef\figin##1{\vskip2in}\gdef\figins##1{\hskip.5in}
\else\message{(FIGURES WILL BE INCLUDED)}%
\gdef\figin##1{##1}\gdef\figins##1{##1}\fi}
\def\DefWarn#1{}
\def\figinsert{\goodbreak\midinsert}
\def\ifig#1#2#3{\DefWarn#1\xdef#1{fig.~\the\figno}
\writedef{#1\leftbracket fig.\noexpand~\the\figno}%
\figinsert\figin{\centerline{#3}}\medskip\centerline{\vbox{\baselineskip12pt \advance\hsize by
-1truein\noindent\footnotefont{\bf Fig.~\the\figno:\ } \it#2}}
\bigskip\endinsert\global\advance\figno by1}

\lref\HWZ{
  K.~A.~Intriligator,
  ``Anomaly matching and a Hopf-Wess-Zumino term in 6d, N=(2,0) field theories,''
Nucl.\ Phys.\ B {\bf 581}, 257 (2000).
[hep-th/0001205].
}
\lref\YiBZ{
  P.~Yi,
  ``Anomaly of (2,0) theories,''
Phys.\ Rev.\ D {\bf 64}, 106006 (2001).
[hep-th/0106165].
}
\lref\HoweUE{
  P.~S.~Howe, N.~D.~Lambert and P.~C.~West,
  ``The Selfdual string soliton,''
Nucl.\ Phys.\ B {\bf 515}, 203 (1998).
[hep-th/9709014].
}
\lref\BermanEW{
  D.~S.~Berman and J.~A.~Harvey,
  ``The Self-dual string and anomalies in the M5-brane,''
JHEP {\bf 0411}, 015 (2004).
[hep-th/0408198].
}
\lref\KimWC{
  H.~Kim and P.~Yi,
  ``D-brane anomaly inflow revisited,''
JHEP {\bf 1202}, 012 (2012).
[arXiv:1201.0762 [hep-th]].
}
\lref\WittenTW{
  E.~Witten,
  ``Global Aspects of Current Algebra,''
Nucl.\ Phys.\ B {\bf 223}, 422 (1983)..
}
\lref\SeibergDR{
  N.~Seiberg and W.~Taylor,
  ``Charge Lattices and Consistency of 6D Supergravity,''
JHEP {\bf 1106}, 001 (2011).
[arXiv:1103.0019 [hep-th]].
}
\lref\SeibergQX{
  N.~Seiberg,
  ``Nontrivial fixed points of the renormalization group in six-dimensions,''
Phys.\ Lett.\ B {\bf 390}, 169 (1997).
[hep-th/9609161].
}
\lref\HarveyBX{
  J.~A.~Harvey, R.~Minasian and G.~W.~Moore,
  ``NonAbelian tensor multiplet anomalies,''
JHEP {\bf 9809}, 004 (1998).
[hep-th/9808060].
}
\lref\ParkJI{
  D.~S.~Park,
  ``Anomaly Equations and Intersection Theory,''
JHEP {\bf 1201}, 093 (2012).
[arXiv:1111.2351 [hep-th]].
}
\lref\GreenBX{
  M.~B.~Green, J.~H.~Schwarz and P.~C.~West,
  ``Anomaly Free Chiral Theories in Six-Dimensions,''
Nucl.\ Phys.\ B {\bf 254}, 327 (1985).
}
\lref\SeibergVS{
  N.~Seiberg and E.~Witten,
  ``Comments on string dynamics in six-dimensions,''
Nucl.\ Phys.\ B {\bf 471}, 121 (1996).
[hep-th/9603003].
}
\lref\DeserMZ{
  S.~Deser, A.~Gomberoff, M.~Henneaux and C.~Teitelboim,
  ``Duality, selfduality, sources and charge quantization in Abelian N form theories,''
Phys.\ Lett.\ B {\bf 400}, 80 (1997).
[hep-th/9702184].
}
\lref\Eanomaly{
  K.~Ohmori, H.~Shimizu and Y.~Tachikawa,
  ``Anomaly polynomial of E-string theories,''
[arXiv:1404.3887 [hep-th]].
}
\lref\HenningsonDH{
  M.~Henningson,
  ``Self-dual strings in six dimensions: Anomalies, the ADE-classification, and the world-sheet WZW-model,''
Commun.\ Math.\ Phys.\  {\bf 257}, 291 (2005).
[hep-th/0405056].
}
\lref\IntriligatorEX{
  K.~A.~Intriligator and N.~Seiberg,
  ``Mirror symmetry in three-dimensional gauge theories,''
Phys.\ Lett.\ B {\bf 387}, 513 (1996).
[hep-th/9607207].
}
\lref\WittenGX{
  E.~Witten,
  ``Small instantons in string theory,''
Nucl.\ Phys.\ B {\bf 460}, 541 (1996).
[hep-th/9511030].
}

\lref\GanorMU{
  O.~J.~Ganor and A.~Hanany,
  ``Small E(8) instantons and tensionless noncritical strings,''
Nucl.\ Phys.\ B {\bf 474}, 122 (1996).
[hep-th/9602120].
}
\lref\MorrisonXF{
  D.~R.~Morrison and N.~Seiberg,
  ``Extremal transitions and five-dimensional supersymmetric field theories,''
Nucl.\ Phys.\ B {\bf 483}, 229 (1997).
[hep-th/9609070].
}
\lref\IntriligatorPQ{
  K.~A.~Intriligator, D.~R.~Morrison and N.~Seiberg,
  ``Five-dimensional supersymmetric gauge theories and degenerations of Calabi-Yau spaces,''
Nucl.\ Phys.\ B {\bf 497}, 56 (1997).
[hep-th/9702198].
}
\lref\BershadskySB{
  M.~Bershadsky and C.~Vafa,
  ``Global anomalies and geometric engineering of critical theories in six-dimensions,''
[hep-th/9703167].
}
\lref\IntriligatorKQ{
  K.~A.~Intriligator,
  ``RG fixed points in six-dimensions via branes at orbifold singularities,''
Nucl.\ Phys.\ B {\bf 496}, 177 (1997).
[hep-th/9702038].
}
\lref\IntriligatorDH{
  K.~A.~Intriligator,
  ``New string theories in six-dimensions via branes at orbifold singularities,''
Adv.\ Theor.\ Math.\ Phys.\  {\bf 1}, 271 (1998).
[hep-th/9708117].
}
\lref\DelZottoHPA{
  M.~Del Zotto, J.~J.~Heckman, A.~Tomasiello and C.~Vafa,
  ``6d Conformal Matter,''
[arXiv:1407.6359 [hep-th]].
}
\lref\AspinwallYE{
  P.~S.~Aspinwall and D.~R.~Morrison,
  ``Point - like instantons on K3 orbifolds,''
Nucl.\ Phys.\ B {\bf 503}, 533 (1997).
[hep-th/9705104].
}

\lref\KumarAE{
  V.~Kumar and W.~Taylor,
  ``A Bound on 6D N=1 supergravities,''
JHEP {\bf 0912}, 050 (2009).
[arXiv:0910.1586 [hep-th]].
}
\lref\ParkWV{
  D.~S.~Park and W.~Taylor,
  ``Constraints on 6D Supergravity Theories with Abelian Gauge Symmetry,''
JHEP {\bf 1201}, 141 (2012).
[arXiv:1110.5916 [hep-th]].
}
\lref\BlumMM{
  J.~D.~Blum and K.~A.~Intriligator,
  ``New phases of string theory and 6-D RG fixed points via branes at orbifold singularities,''
Nucl.\ Phys.\ B {\bf 506}, 199 (1997).
[hep-th/9705044].
}
\lref\AspinwallNK{
  P.~S.~Aspinwall and M.~Gross,
  ``The SO(32) heterotic string on a K3 surface,''
Phys.\ Lett.\ B {\bf 387}, 735 (1996).
[hep-th/9605131].
}
\lref\MorrisonJS{
  D.~R.~Morrison and W.~Taylor,
  ``Toric bases for 6D F-theory models,''
Fortsch.\ Phys.\  {\bf 60}, 1187 (2012).
[arXiv:1204.0283 [hep-th]].
}
\lref\HeckmanPVA{
  J.~J.~Heckman, D.~R.~Morrison and C.~Vafa,
  ``On the Classification of 6D SCFTs and Generalized ADE Orbifolds,''
JHEP {\bf 1405}, 028 (2014).
[arXiv:1312.5746 [hep-th]].
}
\lref\HoravaMA{
  P.~Horava and E.~Witten,
  ``Eleven-dimensional supergravity on a manifold with boundary,''
Nucl.\ Phys.\ B {\bf 475}, 94 (1996).
[hep-th/9603142].
}
\lref\SchwarzZW{
  J.~H.~Schwarz,
  ``Anomaly - free supersymmetric models in six-dimensions,''
Phys.\ Lett.\ B {\bf 371}, 223 (1996).
[hep-th/9512053].
}
\lref\SagnottiQW{
  A.~Sagnotti,
  ``A Note on the Green-Schwarz mechanism in open string theories,''
Phys.\ Lett.\ B {\bf 294}, 196 (1992).
[hep-th/9210127].
}
\lref\AlvarezGaumeDR{
  L.~Alvarez-Gaume and P.~H.~Ginsparg,
  ``The Structure of Gauge and Gravitational Anomalies,''
Annals Phys.\  {\bf 161}, 423 (1985), [Erratum-ibid.\  {\bf 171}, 233 (1986)]..
}
\lref\BardeenPM{
  W.~A.~Bardeen and B.~Zumino,
  ``Consistent and Covariant Anomalies in Gauge and Gravitational Theories,''
Nucl.\ Phys.\ B {\bf 244}, 421 (1984).
}
\lref\HarveyIT{
  J.~A.~Harvey,
  ``TASI 2003 lectures on anomalies,''
[hep-th/0509097].
}

\lref\AlvarezGaumeIG{
  L.~Alvarez-Gaume and E.~Witten,
  ``Gravitational Anomalies,''
Nucl.\ Phys.\ B {\bf 234}, 269 (1984).
}
\lref\BernardNR{
  C.~W.~Bernard, N.~H.~Christ, A.~H.~Guth and E.~J.~Weinberg,
  ``Instanton Parameters for Arbitrary Gauge Groups,''
Phys.\ Rev.\ D {\bf 16}, 2967 (1977)..
}
\lref\EguchiJX{
  T.~Eguchi, P.~B.~Gilkey and A.~J.~Hanson,
  ``Gravitation, Gauge Theories and Differential Geometry,''
Phys.\ Rept.\  {\bf 66}, 213 (1980)..
}
\lref\OSTY{
  K.~Ohmori, H.~Shimizu, Y.~Tachikawa and K.~Yonekura,
  ``Anomaly polynomial of general 6d SCFTs,''
[arXiv:1408.5572 [hep-th]].
}
\lref\SadovZM{
  V.~Sadov,
  ``Generalized Green-Schwarz mechanism in F theory,''
Phys.\ Lett.\ B {\bf 388}, 45 (1996).
[hep-th/9606008].
}
\lref\GanorVE{
  O.~Ganor and L.~Motl,
  ``Equations of the (2,0) theory and knitted five-branes,''
JHEP {\bf 9805}, 009 (1998).
[hep-th/9803108].
}
\lref\BolognesiRQ{
  S.~Bolognesi and K.~Lee,
  ``1/4 BPS String Junctions and $N^3$ Problem in 6-dim (2,0) Superconformal Theories,''
Phys.\ Rev.\ D {\bf 84}, 126018 (2011).
[arXiv:1105.5073 [hep-th]].
}
\lref\HollowoodCG{
  T.~J.~Hollowood and P.~Mansfield,
  ``Rational Conformal Field Theories At, and Away From, Criticality as Toda Field Theories,''
Phys.\ Lett.\ B {\bf 226}, 73 (1989)..
}

\newbox\tmpbox\setbox\tmpbox\hbox{\abstractfont }
\Title{\vbox{\baselineskip12pt \hbox{UCSD-PTH-14-08}}}
{\vbox{\centerline{6d, ${\cal N}=(1,0)$ Coulomb Branch Anomaly Matching}}}
\smallskip
\centerline{Kenneth Intriligator}
\smallskip
\bigskip
\centerline{{\it Department of Physics, University of
California, San Diego, La Jolla, CA 92093 USA}}

\bigskip
\vskip 1cm

\noindent  6d QFTs are constrained by the analog of 't Hooft anomaly matching: all anomalies for global symmetries and metric backgrounds are constants of RG flows, and for all vacua in moduli spaces. We discuss an anomaly matching mechanism for 6d ${\cal N}=(1,0)$ theories on their Coulomb branch.  It is a global symmetry analog of Green-Schwarz-West-Sagnotti anomaly cancellation, and requires  the apparent anomaly mismatch to be a perfect square, $\Delta I_8=\half X_4^2$.   Then $\Delta I_8$ is cancelled by making $X_4$ an  electric / magnetic source for the tensor multiplet, so background gauge field instantons yield charged strings. This requires the coefficients in $X_4$ to be integrally quantized.  We illustrate this for ${\cal N}=(2,0)$ theories.  We also consider the ${\cal N}=(1,0)$ SCFTs from $N$ small $E_8$ instantons, verifying that the recent result for its anomaly polynomial fits with the anomaly matching mechanism. 

\bigskip

\Date{August 2014}

\newsec{Introduction}
Brane constructions in a decoupling limit \SeibergVS\ led to the idea that there are local, interacting, 6d QFTs \SeibergQX. These theories cannot be formulated in any known, conventional lagrangian description, because they contain {\it interacting} two-form gauge fields, with self-dual field strength: the challenge is that the charged objects would be string-like, with self-dual electric-magnetic charges. Examples include the 6d ${\cal N}=(2,0)$ theories, the ${\cal N}=(1,0)$ include the theory of $N$ small $E_8$ instantons\foot{Dimensionally reducing the small $E_8$ instanton theory to $d<6$ gives theories that can be related to more conventional QFTs, e.g. \refs{\IntriligatorEX, \MorrisonXF, \IntriligatorPQ}.}\refs{\WittenGX, \GanorMU, \SeibergVS} and many others, obtained from decoupling limits of string, brane, M-theory, or F-theory  constructions, see e.g. \refs{\AspinwallNK\IntriligatorKQ\BlumMM\AspinwallYE \IntriligatorDH \MorrisonJS \HeckmanPVA-\DelZottoHPA}. 

6d QFTs have chiral matter, so anomalies provide a useful handle. Gauge anomaly cancellation highly constrains the matter content \refs{\SeibergQX, \IntriligatorKQ, \IntriligatorDH,  \SchwarzZW\BershadskySB  \KumarAE- \ParkWV}.  The analog of 't Hooft anomalies, for global symmetries, usefully constrains the low-energy theory: these anomalies must be constant along RG flows, and on the vacuum manifold, even if the symmetry is spontaneously broken.  In the broken case, as in 4d \WittenTW, anomaly matching can require certain WZW-type low-energy interactions, to cancel apparent anomaly mismatches. This was discussed for 6d theories in \HWZ, and applied to the case of 
${\cal N}=(2,0)$ theories on the Coulomb branch.  We here apply analogous considerations to ${\cal N}=(1,0)$ theories.  
 
Consider a  6d, ${\cal N}=(1,0)$ theory with a Coulomb branch moduli space of vacua, associated with $\ev{\phi}$ for the real scalar(s) of tensor multiplets.  Let 
${\cal S}^{\rm origin}$ denote the low-energy theory at $\ev{\phi}=0$.  Moving to $\ev{\phi}\neq 0$, the theory reduces at low-energy as 
\eqn\breakgen{{\cal S}^{\rm origin}\qquad \to \qquad {\cal S}^{\rm away}+{\cal S}[U(1)]+\hbox{\it anomaly matching terms}.}
Here ${\cal S}[U(1)]$ denotes a 6d ${\cal N}=(1,0)$ tensor multiplet\foot{The notation is because it reduces, on an $S^1$, to a 5d ${\cal N}=1$, $U(1)$ vector multiplet.}: i.e. a real scalar, $\phi$, a 2-form gauge field $B$ with self-dual field strength $H$, and fermion superparters.   
The non-compact, real $\phi $ is the dilaton of spontaneously-broken conformal symmetry.   The details of the $\rightarrow$ step in \breakgen\ involve integrating out poorly understood interactions, including effective strings coupling ${\cal S}^{away}$ to the $B$ in ${\cal S}[U(1)]$), with string tension $\sim \ev{\phi}\neq 0$.  The {\it anomaly matching terms} in \breakgen\ are non-decoupling effects, regardless of how large $\phi$ is.  Such anomaly-matching-derived interactions can provide useful clues about the dynamics.

Let $I_8^{\rm origin}$ be the anomaly polynomial 8-form of ${\cal S}^{\rm origin}$, and  $I_8^{\rm away, naive}$ that of ${\cal S}^{\rm away}+{\cal S}[U(1)]$.  Any apparent mismatch, $\Delta I_8\equiv I_8^{\rm origin}-I_8^{\rm naive, away}$ must be balanced by some remaining interactions in the low-energy theory.  We here discuss an anomaly matching mechanism, which cancels $\Delta I_8$ provided that it is a perfect square:
\eqn\Idiff{I_8^{\rm origin}-I_8^{\rm away, naive}\equiv \Delta I _8= \half X_4\wedge X_4.}  More generally, with multiple tensors, we need
\eqn\hisource{\Delta I_8=\half \Omega _{IJ}X_4^I \wedge X_4^J \equiv \half \vec X \wedge \cdot \vec X,}
where the $I$ index runs over the tensor multiplets, and $\Omega _{IJ}$ is a positive definite, symmetric metric on the space of tensor multiplets, which is implicit in the $\wedge \cdot$ product in \hisource.

The mechanism is analogous to that of \refs{\GreenBX, \SagnottiQW} for canceling anomalies of local symmetries.  A reducible gauge anomaly $I_8$ can be cancelled via an additional tensor multiplet contribution $\Delta I_8$ of the form\foot{In \refs{\GreenBX, \SagnottiQW}, the $H^I$ also includes the tensor from the gravity multiplet, which has 
 opposite chirality from those of the matter multiplets, and correspondingly enters into $\Omega _{IJ}$ with opposite signature \SagnottiQW.  Here we decouple gravity, so $\Omega _{IJ}$ 
has a definite signature. We take it to be positive.} \hisource.  This is achieved by making $X_4^I$ into electric / magnetic sources for the tensor multiplet field strengths $H^I$.   Our sign conventions\foot{We take matter fermions to contribute positively to $I_8$,  while gauginos contribute negatively.  Then the positive $\Delta I_8$ \hisource\ from  tensor multiplets can e.g. cancel a negative $I_8$ gauge anomaly.} are such that $\Omega _{IJ}$ is positive definite.  The full theory is then gauge anomaly free if  $I_8+\Delta I_8=0$.

We apply a similar mechanism to global symmetries; rather than canceling an unwanted $I_8$ of opposite sign, here the tensor multiplet's $\Delta I_8$ provides the 't Hooft anomaly matching deficit. 
This is achieved by making $\vec X_4$ (the $\vec{\cdot }$ is shorthand for multiple tensors, i.e. the $I$ index in \hisource) act as electric / magnetic sources for the tensor multiplets, so 
\eqn\anomc{S_{eff, low}\supset - \int _{M_6} \vec B_2 \wedge  \cdot  \vec X_4,}
and the magnetic dual effect (see section 2 for details)
\eqn\dHis{d\vec H=\half 2\pi \vec X_4, \qquad\hbox{so}\qquad}
\eqn\dH{\vec H_3=d\vec B_2+\pi \vec X_3^{(0)}, \qquad\hbox{ where}\qquad \vec X_4=d\vec X_3^{(0)}.}
Because  $\vec X_3^{(0)}$ in \dH\ is not invariant under global symmetry background gauge transformations, $\vec B_2$ must also correspondingly transform, such that  $H$ is invariant, $\delta H=0$:
\eqn\Bshift{\delta \vec B_2=-\pi \vec X_2^{(1)},\qquad\hbox{where}\qquad \delta \vec X_3^{(0)}\equiv d \vec X_2^{(1)}.}
 Then variation of \anomc\ will compensate for the apparent discrepancy from \Idiff. 

Because $\vec B_2$ has quantized charges, the coefficients in $\vec X_4$ must be correspondingly appropriately quantized.  The general $\vec X_4$ can be expanded in characteristic classes
\eqn\xfour{\vec X_4=\vec n_{grav} {p_1(T)\over 4}+\vec n_{SU(2)_R} c_2(F_{SU(2)_R})+\sum _i \vec n_i c_2(F_i),}
$p_1(T)$ is the Pontryagin class for the rigid, background spacetime curvature, $p_1(T)\equiv\half  \tr (R/2\pi)^2$, $c_2(R)$ and $c_2(F_i)$ are Chern classes of the  $SU(2)_R$ and $F_i$ flavor symmetry background field strengths. The Chern classes $c_2(R)$ and $c_2(F_i)$
will here always be normalized to integrate to one for the minimal associated instanton configuration in the background gauge fields; as we will discuss, the corresponding statement for $p_1(T)/4$ is less clear.  Such background gauge field instanton configurations are codimension 4 strings\foot{It would be interesting to consider the codimension 4 BPS soliton string configurations \HoweUE, and the analog of 't Hooft anomaly matching for the 2d string worldsheet \refs{\HenningsonDH, \BermanEW}.}, with $\vec H$ charge given by  $\vec n_{SU(2)_R}$ or $\vec n_i$ (the $i$ index runs over all global symmetries). These charges must reside in an integral lattice, so there is a quantization condition
\eqn\nquant{\vec n_{SU(2)_R}\in \vec{\Z}, \qquad \hbox{and} \qquad \vec n_i \in \vec{\Z}.}
We expect that $\vec n_{grav}$ in \xfour\ is also quantized, but are uncertain about the normalization. 

Note also that the susy completion of \anomc\ will give terms ${\cal L}_{eff}\sim -\phi F_{\mu \nu}F^{\mu \nu}$, as in \SeibergQX, now coupling the real scalar $\phi$ of the tensor multiplets to the background field strengths. 

The outline is as follows.  In section 2, we elaborate on the above anomaly matching mechanism.  
In section 3, we discuss the ${\cal N}=(2,0)$ theories, from a ${\cal N}=(1,0)$ perspective. 
In section 4, we review the 6d ${\cal N}=(1,0)$ theories associated with small $E_8$ instantons, and their recently-obtained anomaly polynomial \Eanomaly.  In section 5, apply the anomaly matching mechanism to the  small $E_8$ instanton theory on its Coulomb branch.  
\medskip 
\noindent
{\bf Note added:}  Just prior to posting this paper, the outstanding paper \OSTY\ appeared. It uses essentially the same kind of anomaly matching mechanism as discussed here, to derive new results for anomaly polynomials for many classes of ${\cal N}=(1,0)$ theories.  
 
 \newsec{6d 't Hooft anomalies, and a new mechanism for their matching}
 
By the descent procedure \refs{\AlvarezGaumeIG\AlvarezGaumeDR\BardeenPM-\HarveyIT}, the anomalous variation of the effective action of a 6d theory is given in terms of the anomaly polynomial\foot{The normalization of $I_{d+2}$ is such that a Weyl fermion contributes $\widehat A(T)\tr e^{iF/2\pi}|_{d+2}$ .} 8-form $I_8$:
\eqn\anomdesc{\delta S_{eff}=2\pi \int _{M_6}I_6^{(1)}, \qquad\hbox{where}\quad I_8=dI_7^{(0)}, \qquad \hbox{and} \qquad \delta I_7^{(0)}=dI_6^{(1)},}
where $\delta$ denotes the variation, $M_6$ is 6d spacetime\foot{There would be a $(-1)^{d/2}$ factor in  \anomdesc\ in Minkowski $M_d$ with mostly $+$ signature \KimWC; we here use Euclidean signature to avoid writing the $-$ sign.}, the subscript on $X_6^{(1)}$ is the form number, and the superscript the order in the gauge or global symmetry variation parameter.

Now suppose that the theory has a moduli space of vacua, and the theory at the origin has anomaly polynomial $I_8^{\rm origin}$, while the theory away from the origin has a naively different anomaly polynomial $I_8^{\rm away, naive}$.  The naive difference leads to an apparent mismatch
\eqn\anomdiff{\Delta (\delta S_{eff})\equiv \delta S_{eff}^{\rm origin}-\delta S_{eff}^{\rm naive, away}=2\pi \int _{M_6}\Delta I_6^{(1)}, \qquad\hbox{with} \qquad \Delta I_8\equiv I_8^{\rm origin}-I_8^{\rm away, naive}.}
The variation of the low-energy effective action must make up for this difference:
\eqn\anomcomp{\delta S_{eff, low}=2\pi \int _{M_6}\Delta I_6^{(1)}.}

As an example, consider  ${\cal N}=(2,0)$ theories on their Coulomb branch:
\eqn\tcoulomb{{\cal T}[G ]\qquad \to\qquad  {\cal T}[H]\times {\cal T}[U(1)] +\hbox{\it anomaly matching interactions}.}
Here ${\cal T}[G]$ denotes the ${\cal N}=(2,0)$ theory of ADE group type $G$, and ${\cal T}[U(1)]$ denotes a free ${\cal N}=(2,0)$ tensor multiplet.  The global $Sp(2)_R\cong SO(5)_R$ is broken in \tcoulomb, as $SO(5)_R\to SO(4)_R$.  The five real scalars $\phi ^{A=1\dots 5}$ of ${\cal T}[U(1)]$ can be regarded as a radial dilaton mode, for spontaneously broken conformal invariance, and Nambu-Goldstone boson modes $S^4\cong SO(5)_R/SO(4)$.  The $SO(5)_R$ 't Hooft anomaly naively does not match, $\Delta I_8=(c (G) -c(H))p_2(F_{SO(5)_R})/24$, where $p_2(F_{SO(5)_R})$ is the 2nd Pontryagin class of the $SO(5)_R$ background field strength, and the needed term \anomcomp\ comes from \HWZ 
\eqn\hopfwzw{S_{eff, low}\supset 2\pi {c(G)-c(H) \over 6}\int _{M_7} \Omega _3(\phi , A)\wedge d\Omega _3(\phi, A),}
with $d\Omega _3= \phi ^* (\omega _4)$ the volume form on the $S^4$ Nambu-Goldstone manifold, and $\partial M_7=M_6$. 
It was conjectured in  
\HWZ\ that  $c(G)=|G|h_G$, which fits with the $G=SU(N)$ cases \HarveyBX, and also $SO(2N)$ \YiBZ, as derived via M- theory M5 branes and bulk anomaly inflow.

The interaction \hopfwzw\ remains even when the global symmetry background is turned off, $F_{Sp(2)}\to 0$.  This is related to the fact that the 't Hooft anomaly difference, $\Delta I_8 \propto p_2(F_{Sp(2)})$, is irreducible (i.e. it includes $\tr F_{Sp(2)}^4$, not just $(\tr F_{Sp(2)}^2)^2$). This is 
similar to the 4d Wess-Zumino-Witten interaction \WittenTW\ for matching the irreducible 't Hooft anomaly differences of non-Abelian $SU(N\geq 3)$ global symmetries.   Reducible t Hooft anomaly differences, on the other hand, lead to WZW-type interactions that become trivial when the background symmetry gauge fields are set to zero.  That will be the case for the reducible differences \Idiff\ to be discussed here. 

For 't Hooft anomaly discrepancies  of the form \Idiff\ on the Coulomb branch  \breakgen, the  needed compensating variation \anomcomp\ is 
\eqn\anomdiffis{\delta S_{eff, low}=2\pi  \int _{M_6}   \left(\half \vec X_4\wedge \cdot \vec X_4\right)^{(1)}=\pi  \int _{M_6}\vec X_4\wedge \cdot \vec X_2^{(1)},}
where we define $\vec X_3^{(0)}$ and $\vec X_2^{(1)}$ via the usual descent notation, as in \anomdesc :
\eqn\descx{\vec X_4\equiv d\vec X_3^{(0)}, \qquad \delta \vec X_3^{(0)}\equiv d \vec X_2^{(1)}.}
The variation \anomdiffis\ arises from the term \anomc\ in the low-energy effective action. Unlike \hopfwzw, the interaction \anomc\ does not require going to 7d, and it is only non-zero if the global symmetry and metric background fields are non-zero; again, this is because $\Delta I_8$ here is reducible.  Also, the compact global symmetries are unbroken, so there are no Nambu-Goldstone bosons (though $\phi$ is a dilaton).

Note that a self-dual string's charge $\vec Q$ is quantized as \foot{The $\half$ here is from the 6d string's Dirac quantization, $eg=\half ~ 2\pi ~n$, see e.g.  \refs{\DeserMZ \BermanEW-  \KimWC }.}  
\eqn\stringsource{d\vec H=\half ~ 2\pi  ~ \vec Q ~ \delta (\Sigma _2\hookrightarrow M_6), \qquad \vec Q\in \vec{\Z},}
 which expresses the compactness of the gauge invariance of $B$.  More generally, the lattice of allowed dyonic string charges must be self-dual \SeibergDR.  
 The general 4-form $\vec X_4$ in \Idiff\ can be expanded as in \xfour, in terms of properly normalized characteristic classes.  So $\int _{\Sigma _4}c_2(F_G)=1$ for the minimal $SU(2)\subset G$ instanton\foot{I. e. $c_2(F_G)=\lambda (G) ^{-1} ~ \half ~ \tr (F_G/2\pi)^2$, where $\lambda (G)$ 
can be computed as in  e.g. \refs{\BernardNR, \ParkJI}.}, where  $\Sigma _4$ are the 4 Euclidean directions of an instanton configuration, transverse to the $\Sigma _2$ of a string in 6d.   So $c_2(F_{SU(2)_R})$ and $c_2(F_i)$ are smoothed-out versions of the $\delta (\Sigma _2\hookrightarrow M_6)$ in \stringsource, and the $\vec n_{SU(2)_R}$ or $\vec n_i$ in \xfour\ give the $\vec Q$ charge, hence their quantization conditions in \nquant. 

The  quantization of $\vec n_{grav}$ in \xfour\ and \nquant\ is less clear, as it depends on what are the allowed gravitational analog of instanton configurations.  For compact $\Sigma _4$ without boundary, $\int _{\Sigma _4}p_1\in 24\Z$ if $\Sigma _4$ is spin (this follows from the spin $1/2$ index theorem, since $\widehat A=1+p_1/24+\dots$); for compact $\Sigma _4$ that is not necessarily spin,  $\int _{\Sigma _4}p_1\in 3\Z$.   But here we are interested non-compact $\Sigma _4$, or $\Sigma _4$ with boundary, where the index theorems include boundary contributions, $\eta$, and the quantization conditions are weaker, see e.g. \EguchiJX.  The $Q$ contribution from $n_{grav}$ could likewise have boundary contributions.    We will not consider the $n_{grav}$ quantization issue further here.  We will see that the $E_8$ instanton example gives $n_{grav}=1$ with the normalization in \xfour.  

\newsec{${\cal N}=(2,0)$ theories, regarded as a special case of ${\cal N}=(1,0)$}

A ${\cal N}=(2,0)$ theory can be regarded as a special case of a ${\cal N}=(1,0)$ theory, where the global $Sp(1)_R$ enhances to $Sp(2)_R$.  As reviewed around \hopfwzw, the full $Sp(2)_R$ has an irreducible $\Delta I_8$.  But $\Delta I_8$ becomes reducible from the ${\cal N}=(1,0)$ perspective, as we then only turn on background gauge fields in an $SU(2)_L\times SU(2)_R\subset SO(5)_R$, and  then 
\eqn\ptwosq{\Delta I_8={\Delta c \over 24}p_2(F_{SO(5)_R})\to {\Delta c \over 24}\left( c_2(F_{SU(2)_L})-c_2(F_{SU(2)_R})\right) ^2,}
where $\Delta c\equiv c(G)-c(H)$, and we take $c(G)\equiv h_G|G|$.  The $\Delta I_8$ in \ptwosq\ can of course still be matched via \hopfwzw, taking the gauge fields there only in $SU(2)_L\times SU(2)_R$.  

More directly, we can write \ptwosq\ as $\Delta I_8=\half X_4^2$, and match it as in \anomc\ and \dHis.   Superficially, this does not fit with the quantization condition \nquant, since  $\sqrt{\Delta c/12}\notin \Z$; e.g. for $SU(N)\to SU(N-1)\times U(1)$, $\Delta c/3=N(N-1)$, and for $E_8\to E_7\times U(1)$, $\Delta c/6=(29)^2$.   A similar confusion appeared in \HWZ\ (with similar resolution as here), where it was noted that \hopfwzw\ can be obtained 
by taking $d\Omega _3$ to source $H_3$ with coefficient $\alpha _m$ and $\star H_3$ with coefficient $\alpha _e$, see also \GanorVE.  This seemed to require $\Delta c/12=\alpha _e\alpha _m$, with $\alpha _e\neq \alpha _m$, apparently in conflict with self-duality of $H_3$, and unclear quantization of $\alpha _{e, m}$.  

The point is simply that the metric $\Omega _{IJ}$, implicit in \Idiff\ and \hisource, is not $\delta _{IJ}$.  Actually, $\Omega _{IJ}=C_{IJ}^{-1}$, the inverse Cartan matrix of the ADE group $G$ (this is also seen in the related theories of five-branes at orbifold singularities, in \BlumMM).  E.g. for the $G=SU(2)$ theory, $\Omega =\half$, so \ptwosq\ gives $X_4=\sqrt{\Delta c /6}(c_2(F_{SU(2)_L})-c_2(F_{SU(2)_R}))$, which satisfies \nquant\ because here $\Delta c=6$.  More generally, as noted in \BolognesiRQ\ (or \HollowoodCG,  for 2d Toda), the Freudenthal and de Vries strange formula implies that, for $G=A,D,E$, (where $|G|=r_G(h_G+1)$)
\eqn\cissq{{c(G)\over 12}\equiv {h_G|G| \over 12}={1\over 12}f_{abc}f^{abc} ={\bf \vec \rho}\cdot {\bf \vec \rho},}
where $f_{abc}$ are the group structure constants and ${\bf \vec \rho}=\half \sum _{\alpha >0}{\bf \vec \alpha}$ is the Weyl vector. Then \ptwosq, with $\Omega _{IJ}=C_{IJ}^{-1}$ is indeed compatible with the quantization \nquant; it is just obscured a bit by focusing on partial breaking $G\to H\times U(1)$. 

\newsec{Review: the small $E_8$ instanton theory, ${\cal E}_8[N]$, and its anomaly polynomial}

We will illustrate the anomaly matching mechanism for the case ${\cal S}^{\rm origin}={\cal E}_8[N]$, i.e. the theory of $N$ small $E_8$ instantons.    Recall that the case of $N=1$ small $E_8$ instanton has a Higgs branch ${\cal M}_{Higgs}$ that is the $29+1$ hypermultiplet-dimensional moduli space of an $E_8$ instanton.  The $+1$ hypermultiplet here is the translational zero mode of the codimension 4 instanton.  Likewise, for all ${\cal E}_8[N]$, it is convenient to add a free hypermultiplet, for the CM position of the $N$ instantons.  At the origin of the Higgs branch of ${\cal E}_8[N]$, there is an interacting SCFT, with an $N$ real-dimensional, tensor-multiplet, Coulomb branch.   

This structure is evident in the $M$-theory realization, via $N$ coincident M5 branes, which are also coincident with the end-of-the-interval  \HoravaMA\ M9 brane.  The $E_8$ gauge symmetry of the M9 brane becomes the global $E_8$ symmetry of the 6d SCFT in the decoupling limit.  The 6d spacetime directions are  $x^{0, 1,2,3,4,5}$, and the M9 brane is at say $x^{11}=0$.    The Coulomb branch corresponds to moving the M5 branes to $\phi \sim x^{11}\neq 0$ (the Higgs branch corresponds to dissolving the M5s into $E_8$ instantons, necessarily at $x^{11}=0$).    The added free-hypermultiplet corresponds to the CM location of the M5 branes in the $x^{6,7,8,9}$ directions.  By considering anomaly inflow, as in \HarveyBX\ but including the effect of the M9 brane, the anomaly polynomial of this theory was obtained in \Eanomaly\ to be
\eqn\EIis{I_8[{\cal E}_8[N]+{\rm f.h.}]={N^3\over 6}\chi _4^2+{N^2\over 2}\chi _4 I_4+ N(\half I_4^2-{1\over 48}\widehat I_8).}
Here $+{\rm f.h.}$ denotes ``free-hyper:"  The notation in \EIis\ is much as in \Eanomaly\ 
\eqn\chidef{\chi _4\equiv c_2(F_{SU(2)_L})-c_2(F_{SU(2)_R}),}
\eqn\Ifourdef{I_4\equiv -\half c_2(F_{SU(2)_R})-\half c_2(F_{SU(2)_L})+{1\over 4}p_1(T)+c_2(F_{E_8}),}
\eqn\Ihatdef{\widehat I_8\equiv \chi _4^2+p_2(T)-\left(c_2(F_{SU(2)_R})+c_2(F_{SU(2)_L})-\half p_1(T)\right) ^2.}
Our normalization is such that all $\int _{\Sigma _4} c_2(F)=1$ for the minimal instanton configuration. 

In this notation, the anomaly polynomial of the ${\cal N}=(2,0)$ theory of $N$ M5 branes, keeping only $SO(4)\subset SO(5)_R$ background gauge fields, is \HarveyBX
\eqn\ATN{I_8[{\cal T}[SU(N)]]+I_8[{\cal T}[U(1)]]={N^3\over 24}\chi _4^2-{N\over 48}\widehat I_8.}

\newsec{Anomaly matching for ${\cal E}_8[N]$ on its Coulomb branch}

We consider the   ${\cal E}_8[N]$ Coulomb branch associated with giving expectation value to just one of the $N$ tensor multiplets.  In the M5 realization, we move a single M5 to $x^{11}\neq 0$, leaving the  other $N-1$ coincident with the M9 at $x^{11}=0$.  The breaking pattern is 
\eqn\breakone{{\cal E}_8[N]+{\rm f.h.} \qquad \to \qquad {\cal E}_8[N-1]+2({\rm f.h.}) +{\cal S}[U(1)]+\hbox{\it anomaly matching terms}.}
The f.h. on the LHS of \breakone\ is as in \EIis, and goes for the ride, and the other f.h. on the RSH arises in the low-energy theory. 
 The anomaly polynomial $I_8$ of the LHS of \breakone\ is given in \EIis, and likewise for the ${\cal E}_8[N-1]+{\rm f.h.}$ on the RHS, via $N\to N-1$, while 
 that of ${\rm f.h.}+{\cal S}[U(1)]={\cal T}[U(1)]$ is given by setting $N=1$ in \ATN.  Thus the naive difference in anomalies between the LHS and RHS of \breakone\ is
\eqn\idiffe{\eqalign{\Delta I_8&={1\over 24}(4N^3-4(N-1)^3-1)\chi _4^2+\half (N^2-(N-1)^2)\chi _4 I_4+\half I_4^2,\cr
&={1\over 8}(2N-1)^2\chi _4^2+\half (2N-1)\chi _4 I_4+\half I_4^2\cr &=\half \left((N-\half )\chi _4 + I_4\right)^2.}}
It's indeed a perfect square, as required. Moreover,  writing this $X_4$ as in \Idiff, the coefficients are indeed integrally quantized (the $\half $'s in \idiffe\ all cancel or combine to 1)
\eqn\xfoureis{X_4=(N-1)c_2(F_{SU(2)_L})-Nc_2(F_{SU(2)_R})+{1\over 4}p_1(T)+c_2(F_{E_8}),}
i.e. $n_{SU(2)_L}=N-1$, $n_{SU(2)_R}=-N$, and $n_{E_8}=1$: an 
$SU(2)_L$ instanton carries $N-1$ units of $B$-charge, an $SU(2)_R$ instanton has $-N$ units, and an $E_8$ instanton has $1$ unit of $B$-charge.  Also, $n_{grav}=1$ here (recall the discussion at the end of sect. 2). 

Consider e.g. the case of $N=1$ small $E_8$ instanton where the  theory on the RHS of \breakone\ is just the ${\cal N}=(1,0)$ tensor multiplet ${\cal S}[U(1)]$ and two free hypermultiplets.  An $SU(2)_R$ instanton gives a string of $B$-charge $-1$, and an $E_8$ instanton gives one of $B$-charge $+1$.  In the general $N$ case, the ${\cal E}_8[N-1]$ theory at the origin evidently leads to an extra contribution to the $B$-charge of $\pm (N-1)$ for a $SU(2)_{L,R}$ instanton string.

Another breaking pattern is to give non-zero, coincident, expectation values to all $N$ tensor multiplets of the ${\cal E}_8[N]$.  In the M-theory realization, all $N$ of the M5 branes are moved, together, away from the M9 brane.  This gives the breaking pattern
\eqn\breakall{{\cal E}_8[N]+{\rm f.h.} \qquad \to \qquad  {\cal T}[SU(N)]+{\cal T}[U(1)]+\hbox{\it anomaly matching terms},}
where ${\cal T}$ denotes the ${\cal N}=(2,0)$ theories.  The anomaly matching terms are a non-decoupling effect of the M9 brane. The rest of the low-energy theory on the RHS of \breakall\ has an approximate enhancement of $SO(4)_R\to S0(5)_R$, as part of the approximate, accidental enhancement of ${\cal N}=(1,0)\to {\cal N}=(2,0)$; the anomaly matching terms spoil this enhancement.  The anomaly matching needed for \breakall, by  \EIis\ and \ATN, is 
\eqn\anotheri{\Delta I_8={N^3\over 8}\chi _4^2+{N^2\over 2}\chi _4 I_4+{N\over 2}I_4={N\over 2}\left({N\over 2}\chi _4+I_4\right)^2.}
The $N=1$ case of \breakall\ and \anotheri\ coincides with the $N=1$ case of \breakone\ and \idiffe.    More generally, all $N$ tensor multiplets on the RHS of \breakall\ participate in the anomaly matching mechanism, hence the overall $N$ in \anotheri, with an associated lattice of integral charges. 

\bigskip
\noindent {\bf Acknowledgments:}  I would like to thank the organizers (Davide Gaiotto, Jaume Gomis, and Nathan Seiberg) and participants of the Perimeter Institute's {\it Supersymmetric Quantum Field Theories in Five and Six Dimensions} conference in April 2014, for the stimulating conference, the opportunity to give a talk on this work, and following discussions. I would also like to thank Clay C\'{o}rdova and Thomas Dumitrescu for collaboration on a different 6d project. I also gratefully acknowledge the KITP, Santa Barbara, for hospitality and support while part of this work was done,
 in part funded by the National Science Foundation under Grant No. NSF PHY11-25915. This work was also supported by the US Department of Energy under UCSDÕs contract DE-SC0009919, and the Dan Broida Chair.

\listrefs
\end